\documentclass[12pt,times]{elsarticle}
\pdfoutput=1

\usepackage{mathptmx}       
\usepackage{helvet}         
\usepackage{courier}        
\usepackage{type1cm}        
\usepackage{makeidx}         
\usepackage{graphicx}        
\usepackage{multicol}        
\usepackage[bottom]{footmisc}
\usepackage{caption}
\usepackage{subcaption}
\makeindex             

\begin{document}

\title{Trends, Politics, Sentiments, and Misinformation: Understanding People's Reactions to COVID-$19$ During its Early Stages}
\author[1]{Omar Abdel Wahab\corref{c1}}
\ead{omar.abdulwahab@uqo.ca}
\cortext[c1]{Corresponding author}

\author[1]{Ali Mustafa\corref{c1}}
\ead{musa03@uqo.ca}

\author[1]{Andr\'{e} Bertrand Abisseck Bamatakina\corref{c1}}
\ead{abia08@uqo.ca}

\address[1]{Department of Computer Science and Engineering Universit\'{e} du Qu\'{e}bec en Outaouais, 101, Saint-Jean-Bosco, C.P. 1250, succursale Hull, Gatineau, Quebec, J8X 3X7, Canada}
%
%

\begin{abstract}
The sudden outbreak of COVID-$19$ resulted in large volumes of data shared on different social media platforms. Analyzing and visualizing these data is doubtlessly essential to having a deep understanding of the pandemic's impacts on people's lives and their reactions to them. In this work, we conduct a large-scale spatiotemporal data analytic study to understand peoples' reactions to the COVID-$19$ pandemic during its early stages. In particular, we analyze a JSON-based dataset that is collected from news/messages/boards/blogs in English about COVID-19 over a period of $4$ months, for a total of $5.2M$ posts. The data are collected from December $2019$ to March $2020$ from several social media platforms such as Facebook, LinkedIn, Pinterest, StumbleUpon and VK. Our study aims mainly to understand which implications of COVID-$19$ have interested social media users the most and how did they vary over time, the spatiotemporal distribution of misinformation, and the public opinion toward public figures during the pandemic. Our results can be used by many parties (e.g., governments, psychologists, etc.) to make more informative decisions, taking into account the actual interests and opinions of the people.


\end{abstract}

\maketitle
\section{Introduction}
\label{sec:Introduction}
The abrupt outbreak of COVID-$19$ has created a global crisis that had affected not only our physical health but also our mental health and way of living \cite{world2020mental,khanna2020covid}. As a result of the pandemic, social media usage has undeniably gone up. In fact, the stay-at-home orders that followed the rapid outbreak of COVID-$19$ have pushed us to rely more and more on the Internet, not only for entertainment purposes but also to work from home, to pursue our education virtually, and to catch up with family and friends. Moreover, as shopping centres, stores and restaurants closed their doors for months, most of our shopping activities shifted to online. For example, a survey led by the leading media and research organization \textit{Digital Commerce 360} over $4,500$ Influenster (product discovery and reviews platform for consumers) community members in North America reported that social media consumption increased up to $72\%$ and that the posting activities went up to $43\%$ during pandemic times\footnote{https://www.digitalcommerce360.com/2020/09/16/covid-19-is-changing-how-why-and-how-much-were-using-social-media/}.
Add to this the fact that social media had been the number one communication platform for health professionals, governments, universities and organizations to deliver pandemic-related information to the public \cite{featherstone2012provision,hussain2020role}. Thus, it is undeniable that the pandemic and the subsequent nationwide lockdowns have entailed a second to none surge in social media usage across the World.

Consequently, it becomes crucial to perform a deep social media analysis to extract useful insights about the COVID-$19$ pandemic and peoples' reactions to it. Given that the traditional survey methods are time-consuming and expensive to conduct \cite{valdez2020social}, there is a doubtless need for proactive and timely data analytic studies to understand and respond to the speedily emerging effects of the pandemic on our physical and mental health. Several social media analysis studies \cite{massaad2020social,gonzalez2020social,cinelli2020covid,shahi2021exploratory,kolluri2021coverifi,ceron2021fake,mourad2020critical} have been lately conducted in an attempt to understand the impacts of the COVID-$19$ on people's lives and attitudes. Our work aims to complement these studies by providing a large-scale spatiotemporal on peoples' reactions to the COVID-$19$ pandemic during its early stages. Our work differs from these studies from two perspectives: (1) unlike most of these studies which capitalize on twitter data, we analyze in this work data collected from many social media platforms such as Facebook, LinkedIn, Pinterest, StumbleUpon and VK, some of which haven't been included in earlier studies; (2) we focus our study on the first four months of the pandemic in an attempt to understand the evolution of people's reactions and opinions regarding the pandemic over time. 

\subsection{Contributions} We conduct a large-scale study on a dataset \cite{Geva-Dataset} that contains $5.2M$ posts collected from news/message/boards/blogs about COVID-$19$ over a period of $4$ months (December $2019$ to March $2020$). The goal is to understand how people reacted to the COVID-$19$ pandemic during its early stages and how the pandemic affected peoples' opinions on several matters. To attain this goal, we formulate the following specific research questions that we aim to answer through our study:
\begin{enumerate}
  \item How did the number of COVID-$19$-related postings evolve on social media over time?
  \item Which online Web sites were the most consulted by social media users to get updates on COVID-$19$?
  \item How did the interests of social media users in the ramifications of COVID-$19$ vary across the first four months of the pandemic?
  \item How did the spread of illegitimate information evolve over time?
  \item What countries were the most targeted by the posts shared on social media?
  \item What are the temporal and geographic distributions of illegitimate information?
  \item Which public figures were the most mentioned on social media?
  \item What were the public sentiments toward the most mentioned public figures?
\end{enumerate}

\subsection{Organization}
In Section \ref{Sec:RelatedWork}, we review the related studies and highlight the unique contributions of this work. In Section \ref{sec:Analysis}, we first describe the environment and tools used to conduct our study and then present and discuss the results. Finally, in Section \ref{sec:Conclusion}, we summarize the main findings of the paper.

\section{Related Work}
\label{Sec:RelatedWork}
In \cite{massaad2020social}, the authors aim to examine the volume, content, and geo-spatial distribution of tweets related to telehealth during the COVID-$19$ pandemic. To do so, public data on telehealth in the United States collected from Twitter for the period of March $30$, $2020$ to April $6$, $2020$ have been used. The tweets were analyzed using a mixture of cluster analysis and natural language processing techniques. The study suggests the importance of social media in promoting telehealth-favoring policies to counter mental problems in highly affected areas.

In \cite{gonzalez2020social}, the authors aim to study the impact, advantages and limitation of using social networks during the COVID-$19$ pandemic. The authors concluded that social media is important to foster the dissemination of important information, diagnostics, treatments and follow-up protocols. However, according to the authors also, social media can also be negatively used to spread fake data, pessimist information and myths which could contribute in increasing the depression and anxiety among people.

In \cite{cinelli2020covid}, the authors perform a large-scale analysis of COVID-$19$-related data shared on Instagram, Twitter, Reddit, YouTube and Gab. Particularly, they investigate the engagement of social media users with COVID-$19$ and provide a comparative evaluation on the evolution of the discourse on each social media platform. The main finding of the article is that the interaction patterns of each social media along with the distinctiveness of each platform's audience is a crucial factor in information and misinformation spreading.

In \cite{shahi2021exploratory}, the authors investigate the propagation, authors and content of false information related to COVID-$19$. To do so, they gathered $1500$ fact-checked tweets associated with COVID-$19$ for the period of January to mid-July $2020$, of which $1274$ are false and $226$ are partially false. The study suggests that (1) verified twitter accounts including those of organisations and celebrities contributed in generating or propagation misinformation; (2) tweets with false information often tend to defame legit information on social media and (3) authors of false information use less cautious language, seeking to harm others.

In \cite{kolluri2021coverifi}, the authors develop a Web application, called \textit{CoVerifi}, to asses the credibility of COVID-$19$-related news. \textit{CoVerifi} integrates machine learning \cite{wahab2021federated} and human feedback to evaluate the credibility of the news. It first enables users to give a vote on the content of the news, resulting in a labelled dataset. A Bidirectional Long Short-Term Memory (LSTM) machine learning model is then trained on this dataset to predict future false information.

In \cite{ceron2021fake}, the authors propose a Markov-inspired computational method to characterize topics in tweets within an specific period in Brazil. The proposed solution seeks to address the abuse of social media from three perspectives, which are: (1) providing a better understanding of the fact-checking actions during the pandemic; (2) studying the contradictions between the pandemic and political agendas and (3) detecting false information.

In \cite{mourad2020critical}, the authors conduct a large-scale study on Twitter-generated data that spans over a period of two months. The study concludes that social media have been used to delude users and reorient them to extraneous topics and promote wrongful medical measures and information. On the bright side, the authors noted the importance of credible social media users of different roles (e.g., influencers, content developers, etc.) in the battle against the COVID-$19$ pandemic.

The unique contributions of our work compared to these studies are two-fold: (1) while most studies base their analysis on data collected from twitter, we capitalize in this work on data collected from various social media mediums, i.e., Facebook, LinkedIn, Pinterest, StumbleUpon and VK, where some of these mediums are not considered in the previous studies; (2) our study is based on the first four consecutive months of the pandemic with the goal of understanding the evolution of people's reactions and opinions on matters related the pandemic over time. Thus, the insights extracted from our study are original, given the social media platforms and time interval we consider.

\section{Reactions to COVID-$19$ During its Early Stages: Social Media Analytics}
\label{sec:Analysis}
\subsection{Dataset and Implementation Environment}
Our analysis is done on a JSON-based dataset \cite{Geva-Dataset} which is collected from news/message boards/blogs about COVID-$19$ over a period of $4$ month, for a total of $5.2M$ posts. The time frame of the data is December $2019$ to March $2020$. The posts are in English mentioning at least one of the following: ``Covid'', ``CoronaVirus'' or ``Corona Virus''. To analyze the dataset, we employ the MongoDB\footnote{https://www.mongodb.com/} document-oriented, distributed, JSON-based database platform. More specifically, we write the code in the form of MapReduce queries in MongoDB, which helps us analyze large volumes of data in a distributed way and generate useful aggregated results.

\subsection{Analysis Results}
We explain hereafter the results of our analysis in terms of number of posts related to COVID-$19$ over time; number of published news per Web site, per month; geographic distribution of shared news; geographic and temporal trends in fake news; and opinions about public figures.

\subsubsection{Number of Posts Related to COVID-$19$ Over Time}

\begin{figure*}
	\centering
\scalebox{0.75}{
	\includegraphics[bb= 126 503 486 719]{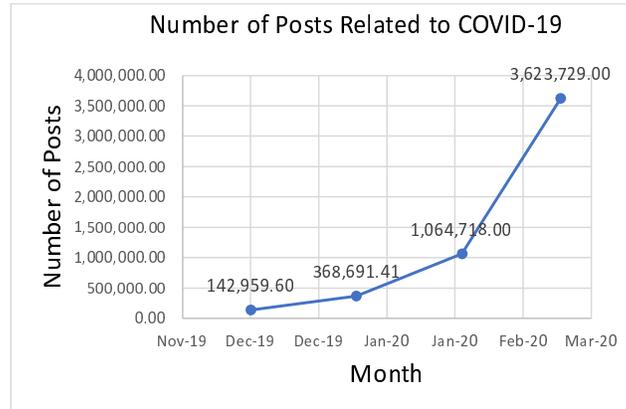}}
	\caption{The number of posts increased exponentially from December $2019$ to March $2020$ with the sharpest increase being in March $2020$.}
	\label{Fig:NumberOfPosts}
\end{figure*}

We study in Fig. \ref{Fig:NumberOfPosts} the evolution of the number of COVID-$19$-related posts on the different studied social media mediums during the first four months of the pandemic. We notice from the figure that the number of posts grew exponentially from December $2019$ to March $2020$. This might be justified by the fact that in December $2019$, the virus was still in its infancy and was somehow limited to China. Starting from January $2020$ when the World Health Organization (WHO) published the first Disease Outbreak News on the COVID-$19$ (January $5$, $2020$) and where the first case of COVID-$19$ was recorded outside of China\footnote{https://www.who.int/news/item/27-04-2020-who-timeline---covid-19}, the number of posts related to COVID-$19$ started to increase exponentially. Yet, it is worth noticing that the sharpest increase in the number of posts was recorded in March $2020$. The reason can be attributed to the fact that during this month, the virus started to spread across the globe and many of the countries started to apply many restrictions such as lockdown and social distancing to contain the spread of COVID-$19$.


\subsubsection{Number of Published News Per Web Site, Per Month}
\begin{figure*}
        \captionsetup[subfigure]{justification=centering}
        \begin{subfigure}{0.55\textwidth}
                \scalebox{0.55}{\includegraphics[bb=120 500 500 719]{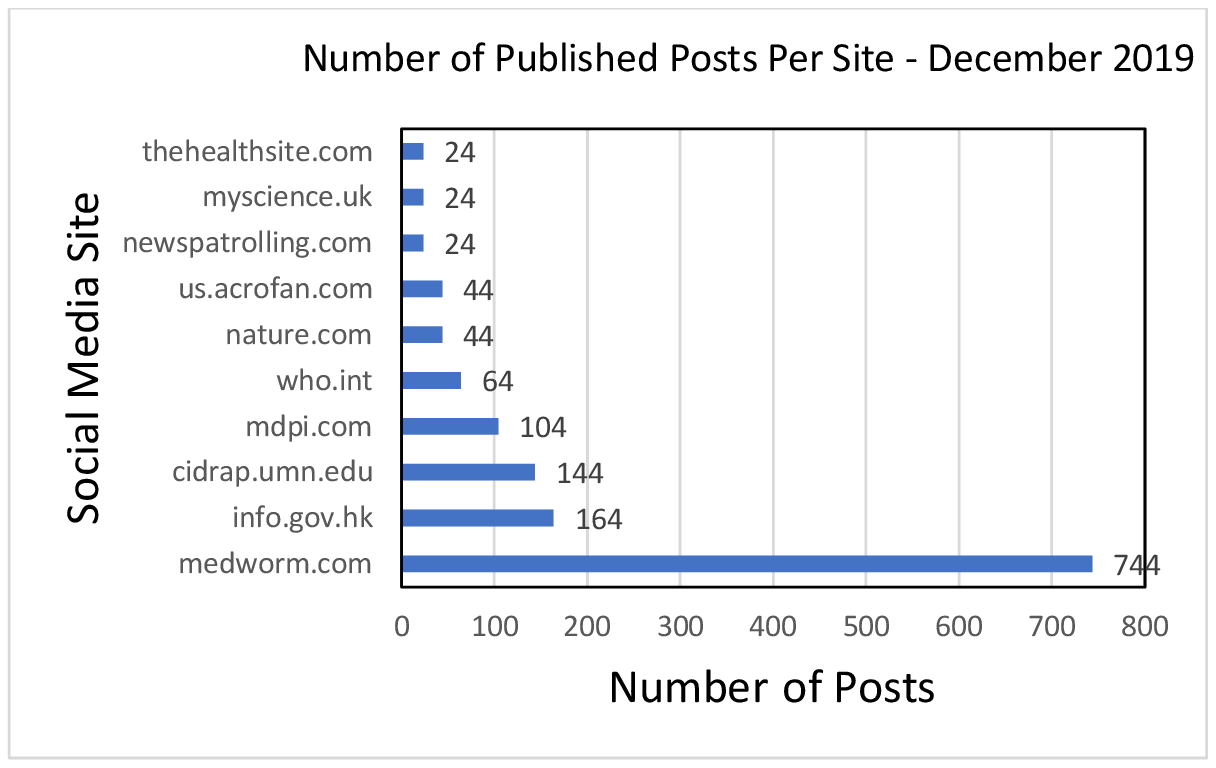}}
                \caption{\footnotesize December $2019$}
                \label{Fig:PostsSiteDecember}
        \end{subfigure}%
        \begin{subfigure}{0.55\textwidth}
                \scalebox{0.55}{\includegraphics[bb=115 500 500 719]{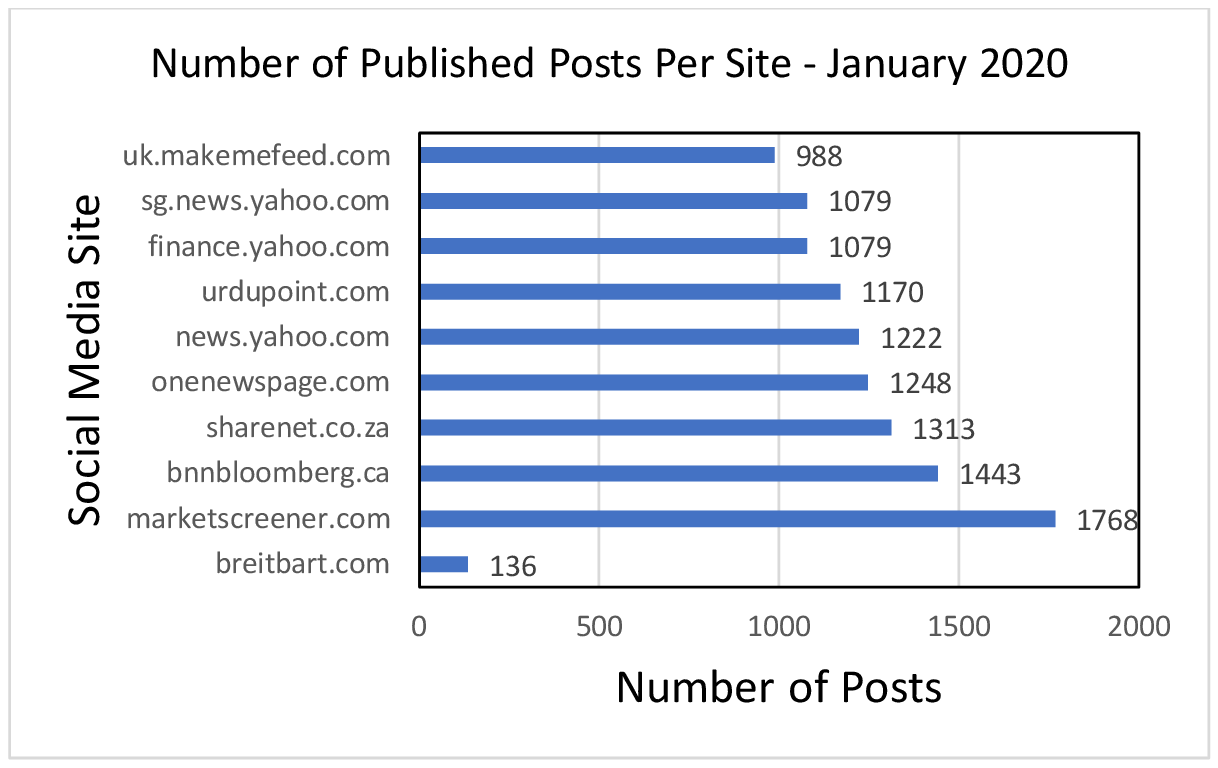}}
                \caption{\footnotesize January $2020$}
                \label{Fig:PostsSiteJanuary}
        \end{subfigure}%

        \begin{subfigure}{0.53\textwidth}
                \scalebox{0.55}{\includegraphics[bb=120 500 500 719]{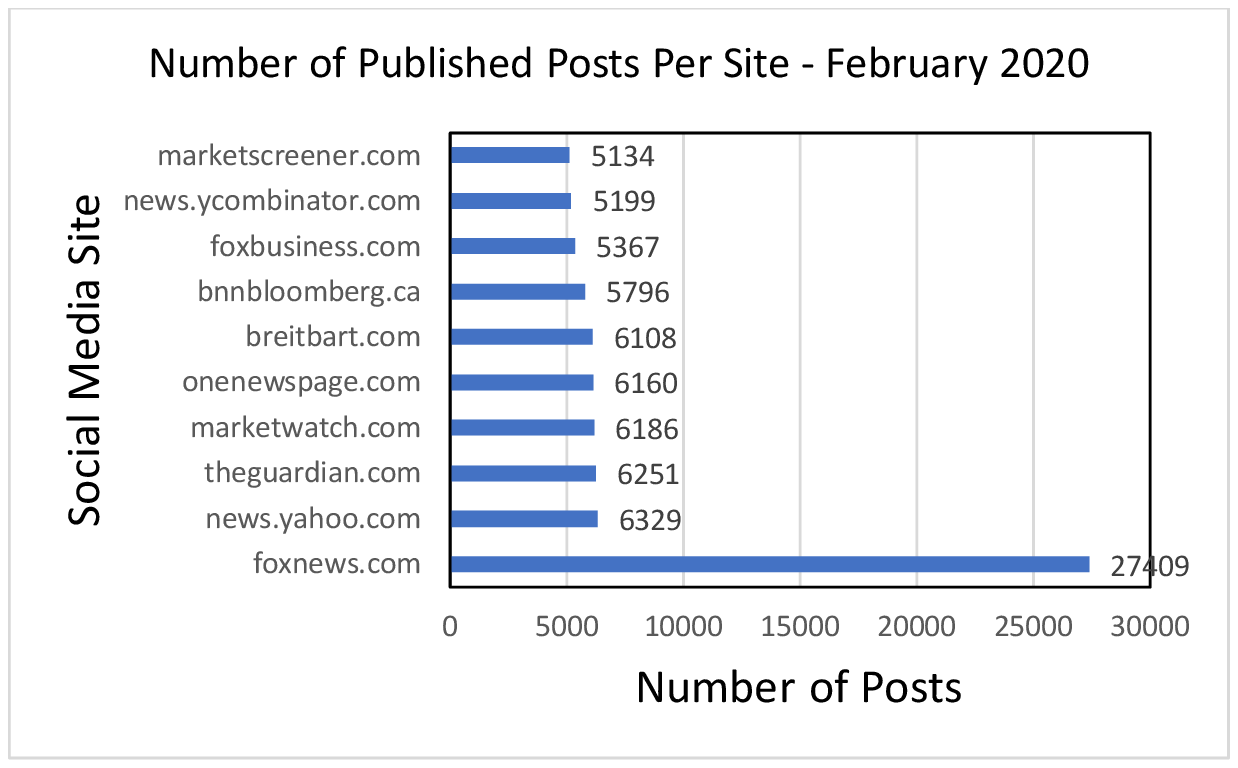}}
                \caption{\footnotesize February $2020$}
                \label{Fig:PostsSiteFebruary}
        \end{subfigure}
        \begin{subfigure}{0.53\textwidth}
                \scalebox{0.55}{\includegraphics[bb=115 500 500 719]{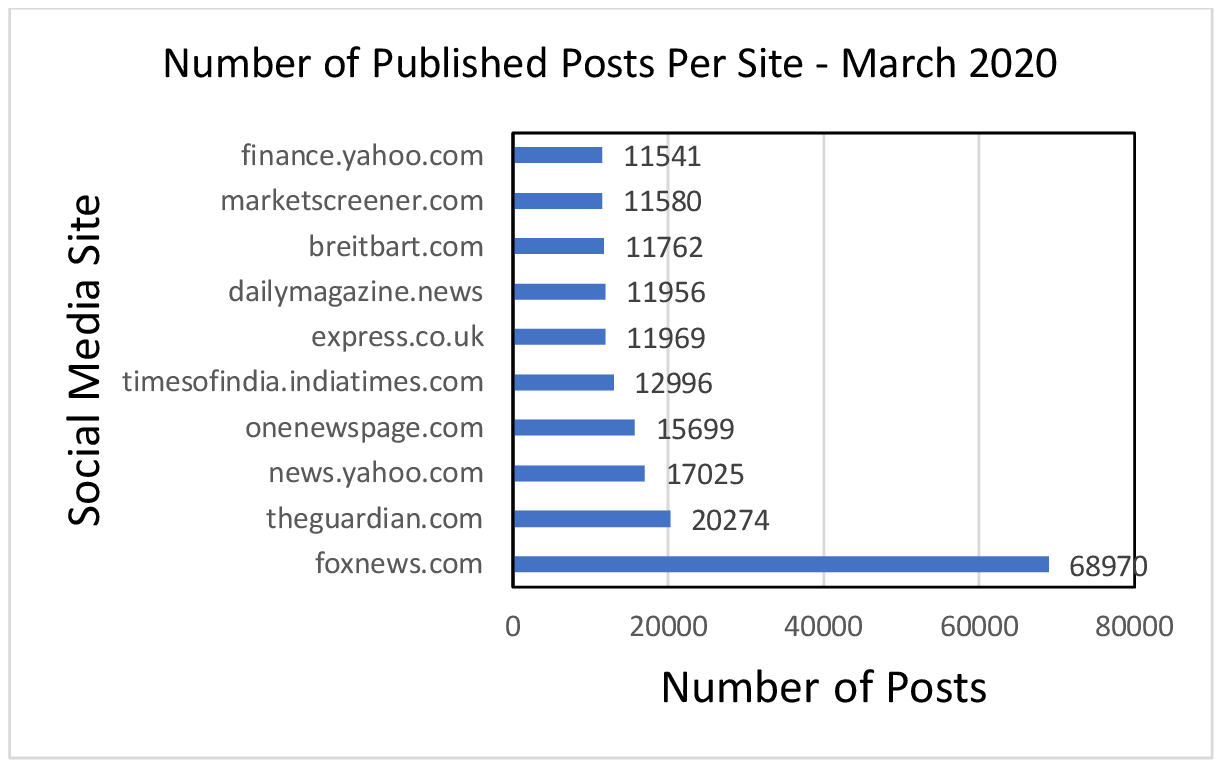}}
                \caption{\footnotesize March $2020$}
                \label{Fig:PostsSiteMarch}
        \end{subfigure}
        \caption{\small The interests of social media users varied significantly across the first four months of the pandemic from being medical-oriented to being market-oriented and then news-oriented.}
        \label{Fig:PostsSiteTime}
\end{figure*}

In Fig. \ref{Fig:PostsSiteTime}, we give a breakdown of the Web sites that were cited on the social media platforms as sources of information for the months of December $2019$ (Fig. \ref{Fig:PostsSiteDecember}), January $2020$ (Fig. \ref{Fig:PostsSiteJanuary}), February $2020$ (Fig. \ref{Fig:PostsSiteFebruary}) and March $2020$ (Fig. \ref{Fig:PostsSiteMarch}). By observing Fig. \ref{Fig:PostsSiteDecember}, we notice that the medical Really Simple Syndication (RSS) feed provider was the most cited Web site in December $2019$ with a big gap vis-\`{a}-vis the other Web sites. This indicates that at that period of the pandemic, the people were mostly interested in learning more about this new generation of viruses from a medical perspective. As for January $2020$, we notice by observing Fig. \ref{Fig:PostsSiteJanuary} that the most cited Web site was \textit{MarketScreener} followed by \textit{BNN Bloomberg}. Knowing that \textit{MarketScreener} is a company that operates as an international stock market and financial news Website and that \textit{BNN Bloomberg} is Canada's Business News Network reporting on finance and markets, we conclude that in the second month of the pandemic, people were more interested in knowing the impacts of the pandemic on the local and global financial markets. On the other hand, we notice from Figures \ref{Fig:PostsSiteFebruary} and \ref{Fig:PostsSiteMarch} that the trend started to change in February and March $2020$ where the most cited Web sites become those that are news-oriented such as \textit{Fox News}, \textit{Yahoo News} and \textit{The Guardian}. This indicates that in this period of time, people started to consult more new-related sites to get news on the emergency measures adopted by the governments and the impacts of the pandemic on the political situation such as the $2020$ United States presidential election \cite{chen2021covid,james2020democratic}.

\subsubsection{Geographic Distribution of Shared News}

\begin{figure*}
        \captionsetup[subfigure]{justification=centering}
        \begin{subfigure}{0.55\textwidth}
                \scalebox{0.50}{\includegraphics[bb=114 509 500 719]{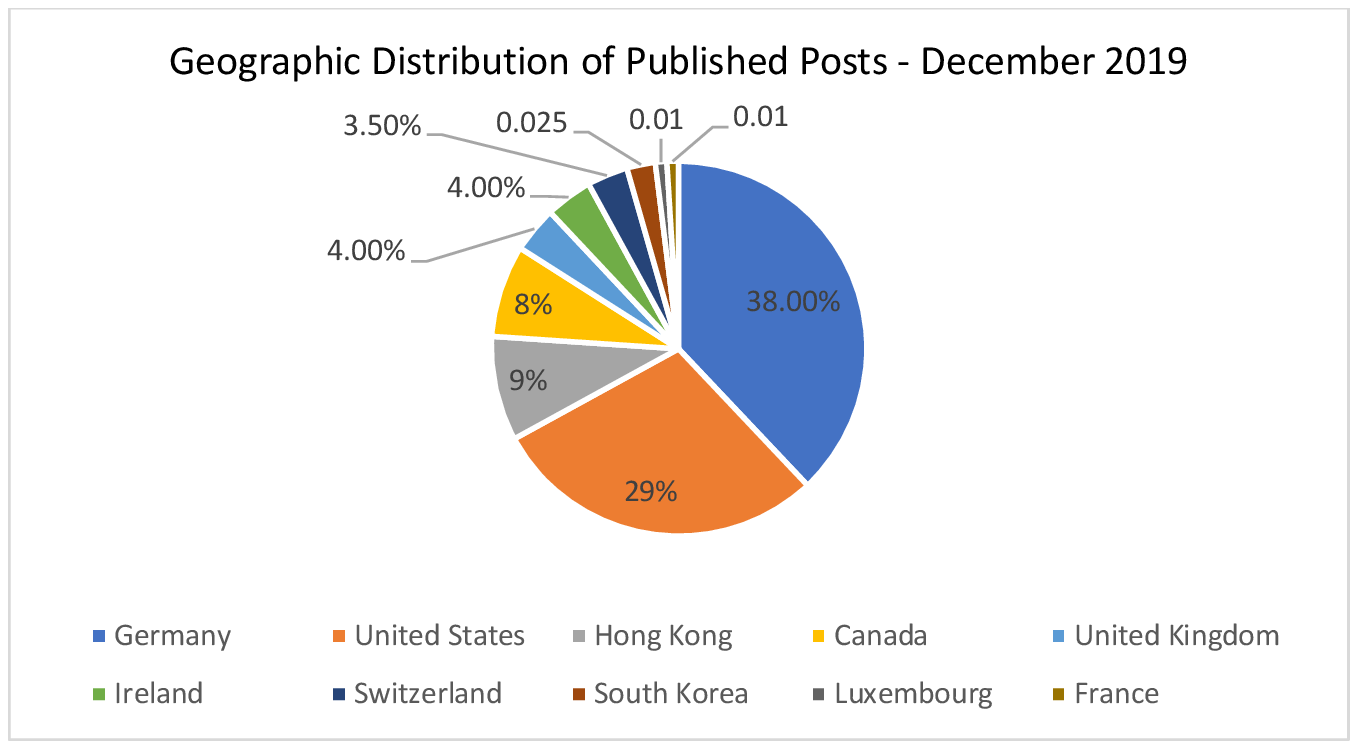}}
                \caption{\footnotesize December $2019$}
                \label{Fig:GeographicNewsDec}
        \end{subfigure}%
        \begin{subfigure}{0.55\textwidth}
                \scalebox{0.48}{\includegraphics[bb=114 509 500 710]{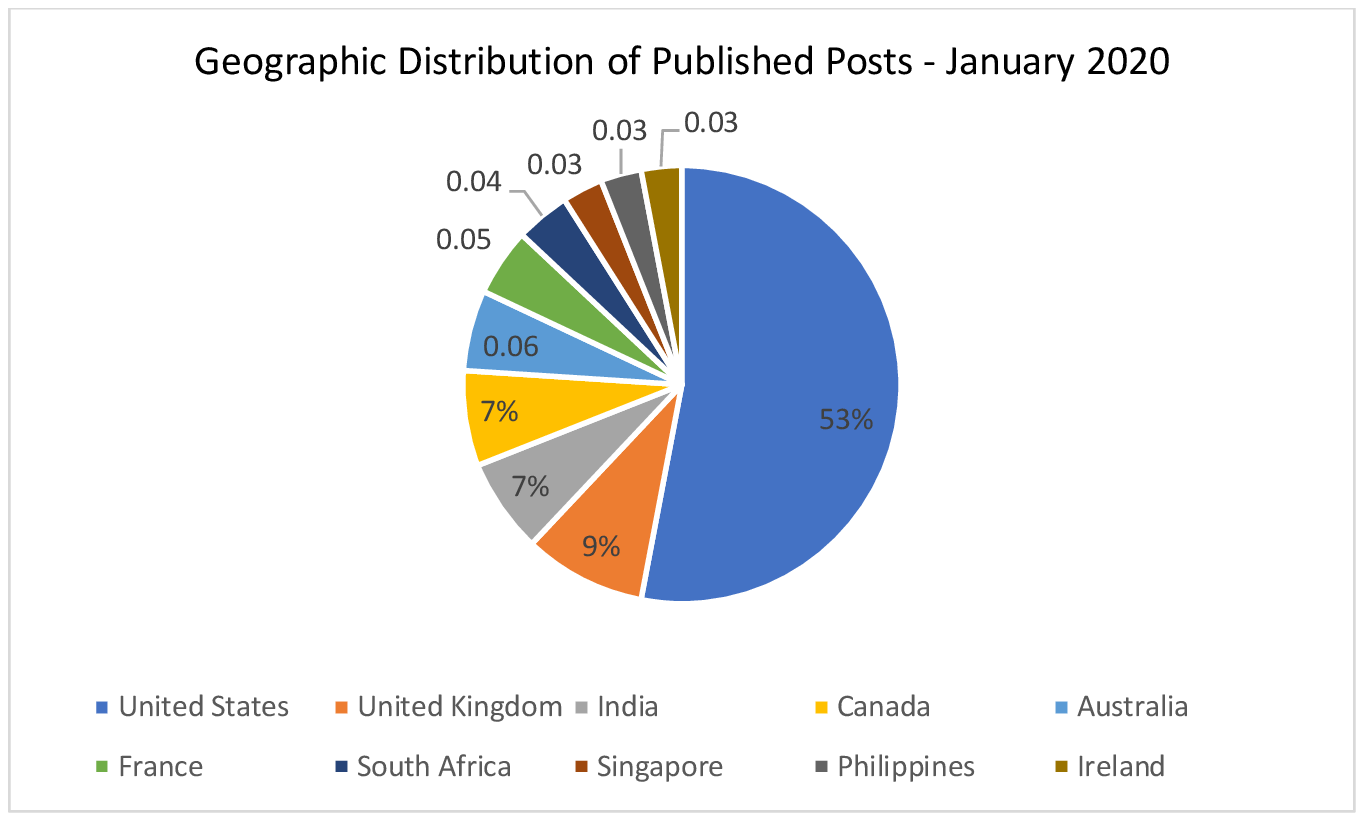}}
                \newline
                \caption{\footnotesize January $2020$}
                \label{Fig:GeographicNewsJan}
        \end{subfigure}%

        \begin{subfigure}{0.55\textwidth}
                \scalebox{0.50}{\includegraphics[bb=114 509 500 725]{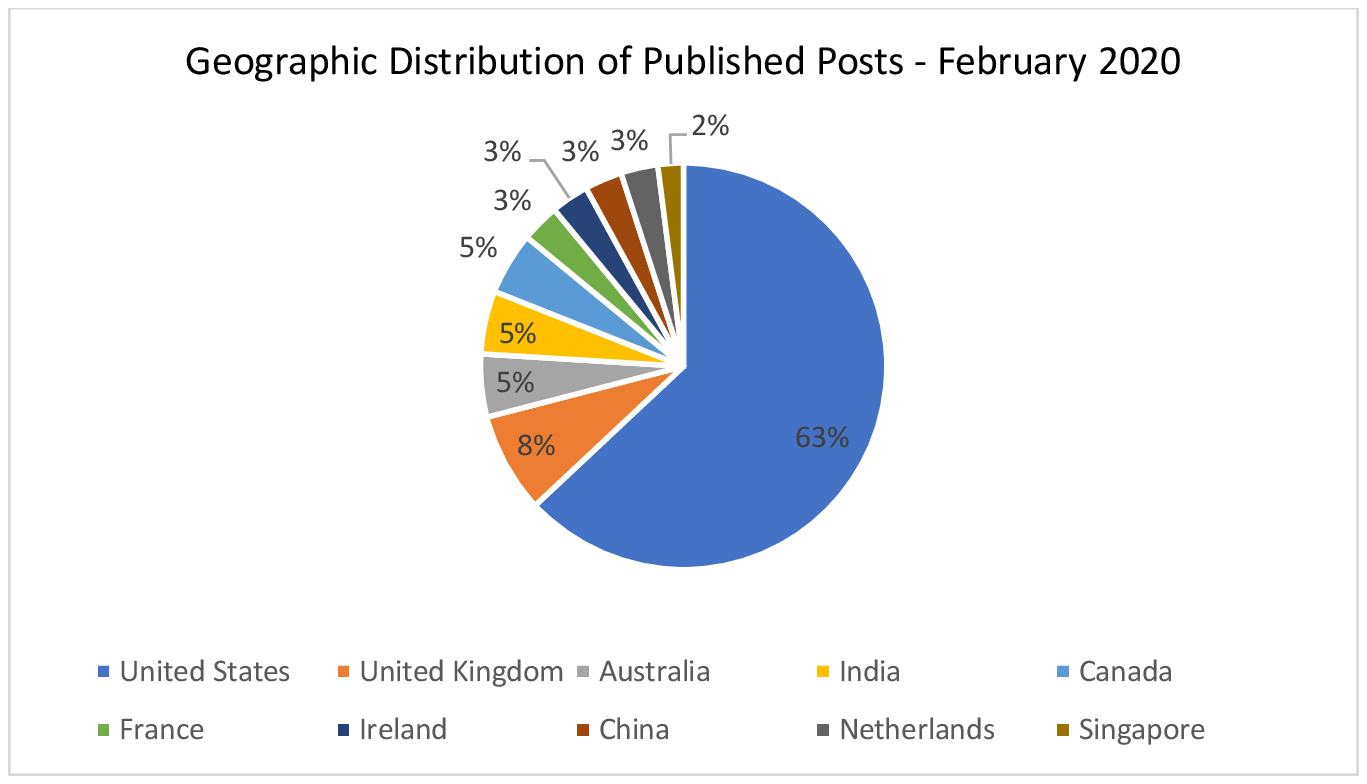}}
                \caption{\footnotesize February $2020$}
                \label{Fig:GeographicNewsFeb}
        \end{subfigure}
        \begin{subfigure}{0.55\textwidth}
                \scalebox{0.50}{\includegraphics[bb=120 509 500 725]{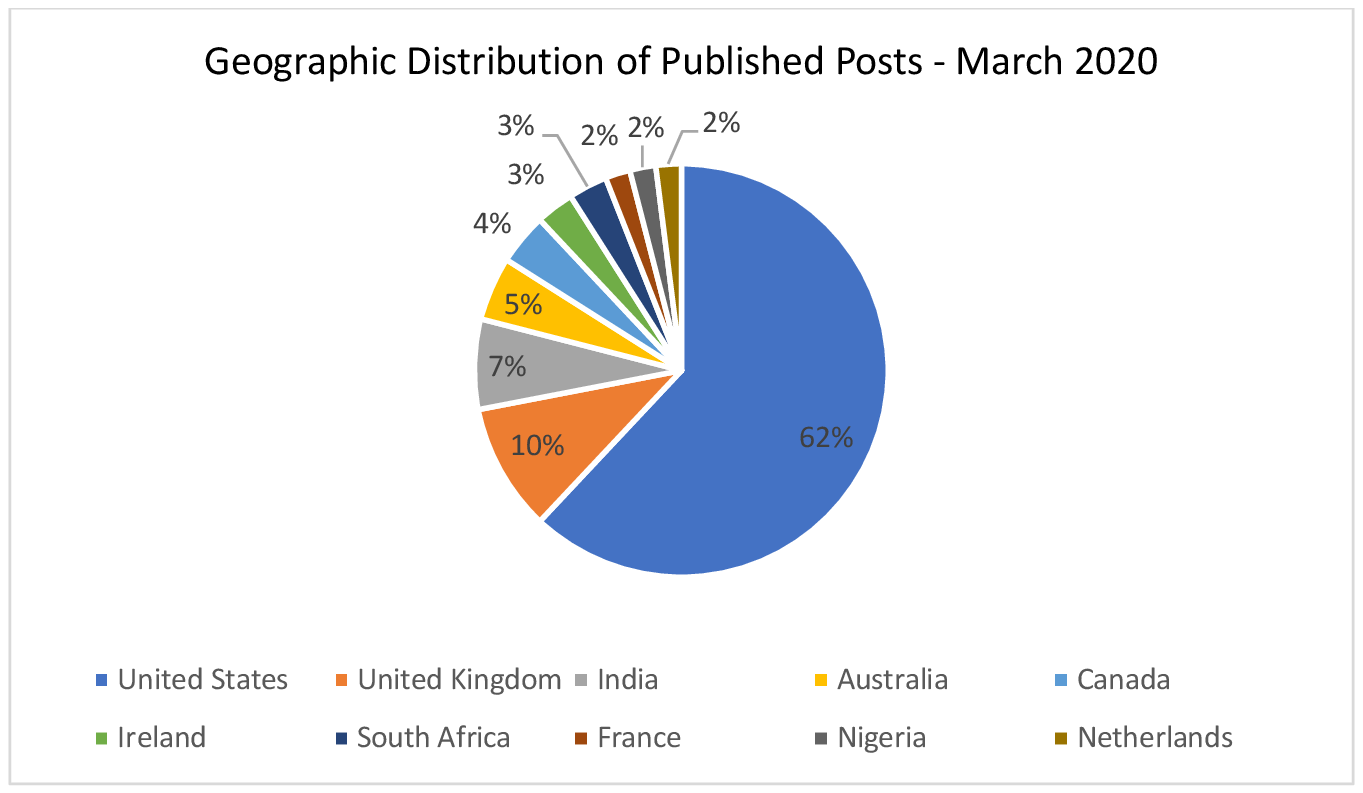}}
                \caption{\footnotesize March $2020$}
                \label{Fig:GeographicNews}
        \end{subfigure}
        \caption{\small Starting from January $2020$, the US accounts for more than the half of the news shared on social media.}\label{Fig:PostsPerSite}
\end{figure*}

We measure in Fig. \ref{Fig:GeographicNews} the geographic distribution of the news shared on social media per month, across the four studied months. Starting with December $2019$ (Fig. \ref{Fig:GeographicNewsDec}), we notice that in this month, Germany accounted for $38\%$ of the news, followed by the United States with a percentage of $29\%$, Hong Kong with a percentage of $9\%$, Canada with a percentage of $8\%$, United Kingdom and Ireland with a percentage of $4\%$, Switzerland  with a percentage of $3.5\%$, South Korea with a percentage of $0.025\%$, and Luxembourg and France with a percentage of $0.01\%$. In January $2020$ (Fig. \ref{Fig:GeographicNewsJan}), we notice that the US accounted for more than the half of the news with a percentage of $53\%$ followed by the United Kingdom with a percentage of $9\%$, with a big noticeable percentage gap between the two countries. We also notice that some new countries started to appear in the shared news such as India, Australia, Singapore, the Philippines and South Africa. In February $2020$ (Fig. \ref{Fig:GeographicNewsFeb}) and March $2020$ (Fig. \ref{Fig:GeographicNewsFeb}), the geographic distribution status quo remains almost the same with the US being in the lead with a percentage of $63\%$ in February $2020$ and a percentage of $62\%$ in March $2020$.

\subsubsection{Geographic and Temporal Trends in Fake News}

\begin{figure*}
	\centering
\scalebox{0.60}{
	\includegraphics[bb= 126 503 486 719]{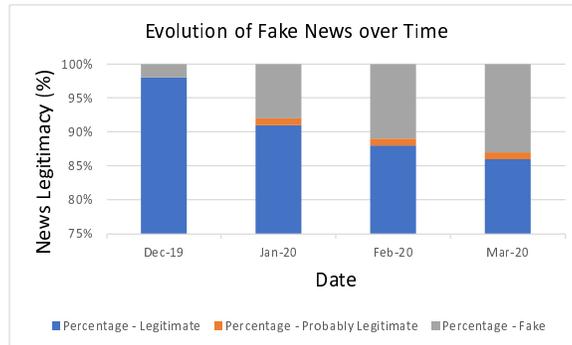}}
	\caption{The amount of fake news increased six times from December $2019$ to March $2020$.}
	\label{Fig:FakeNews}
\end{figure*}

In Fig. \ref{Fig:FakeNews}, we study the evolution of fake news spread across the first four months of the pandemic. The news are classified into three categories, i.e., legitimate, probably legitimate and fake. In the dataset, each shared news is associated with a spam score in the interval $[0,1]$. A Spam Score quantifies the percentage of news with similar features to news that were already classified as illegitimate. To classify the news, we adopt the method proposed by \textit{Link Explorer}\footnote{https://moz.com/help/link-explorer/link-building/spam-score} which is based on the following criteria:
\begin{itemize}
  \item News with a spam score between $1\%$ and $30\%$ are considered legitimate.
  \item News with a spam score between $31\%$ and $60\%$ are considered to be probably legitimate.
  \item  News with a spam score between $61\%$ and $100\%$ are considered illegitimate.
\end{itemize}

By carefully looking at Fig. \ref{Fig:FakeNews}, we notice that the spread of fake news has considerably increased over time. From a percentage of $2\%$ in December $2019$ to a percentage of $13\%$ in March $2020$. Thus, we conclude that the amount of fake news has increased six times in a period of four months.

\begin{figure*}
	\centering
\scalebox{0.60}{
	\includegraphics[bb= 75 475 586 719]{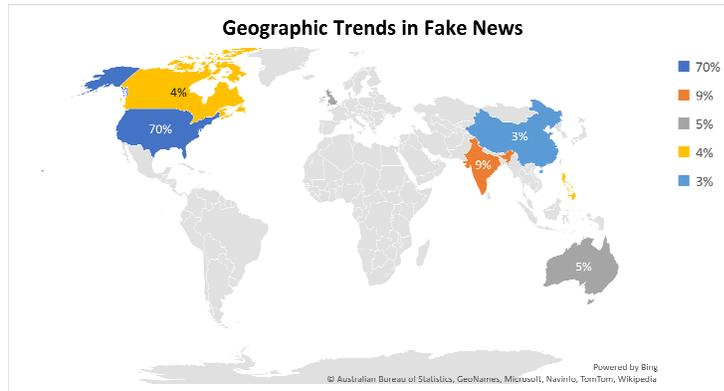}}
	\caption{The United States accounted for $70\%$ of the (English-based) illegitimate news on the studied social media platforms in the period between December $2019$ to March $2020$.}
	\label{Fig:GeographicFake}
\end{figure*}

In Fig. \ref{Fig:GeographicFake}, we study the geographic distribution of fake news. By observing the Figure, we notice that $70\%$ of the fake news came from the United States, followed by $9\%$ from India, $5\%$ from the United Kingdom, $5\%$ from Australia, $4\%$ from the Philippines, $4\%$ from Canada and $3\%$ from China. It is worth mentioning that the fact that the collected news are restricted to the English language only might have influenced the geographic distribution of the news in general, including the illegitimate ones.

\subsubsection{Opinions About Public Figures}
\begin{figure*}
	\centering
\scalebox{0.60}{
	\includegraphics[bb= 176 503 486 719]{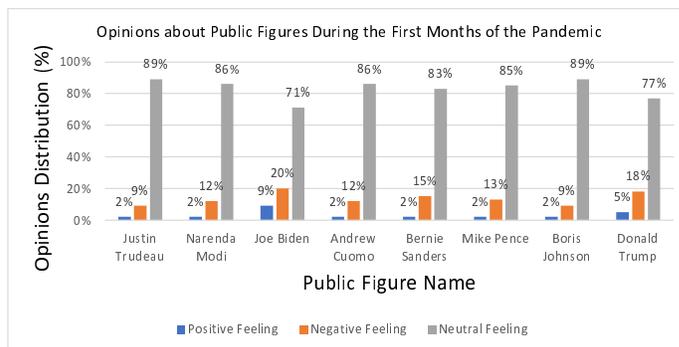}}
	\caption{The most controversial public figures in the period between December $2019$ to March $2020$ were Joe Biden and Donald Trump}
	\label{Fig:SentimentPF}
\end{figure*}

Finally, we identify in Fig. \ref{Fig:SentimentPF}  the public figures that were the most mentioned on social during the first fourth months of the pandemic and provide a detailed breakdown of the overall sentiment of the public towards them. Specifically, the top eight most mentioned public figures on the considered social media platforms in that period were: Justin Trudeau (Prime Minister of Canada), Narenda Modi (Prime Minister of India), Joe Biden (Presidential Candidate at the United States elections during the analyzed period), Andrew Cuomo (New York's Governor), Bernie Sanders (United States Senator), Mike Pence (Vice President of the United States during the analyzed period), Boris Johnson (Prime Minister of the United Kingdom during the analyzed period) and Donald Trump (President of the United States during the analyzed period). To perform the sentiment analysis, we use the \textit{AFINN lexicon} \cite{AFINN} which records over $3,300$+ words with a polarity score (i.e., positive, negative or neutral) associated with each word.

Starting with Justin Trudeau, $89\%$ of the authors were neutral about him, $9\%$ had a negative feeling and $2\%$ had a positive feeling. Moving to Narenda Modi, $86\%$ of the authors were neutral about him, $12\%$ had a negative feeling and $2\%$ had a positive feeling. As for Joe Biden, $71\%$ of the authors were neutral about him, $20\%$ had a negative feeling and $9\%$ had a positive feeling. Concerning Andrew Cuomo, $86\%$ of the authors were neutral about him, $12\%$ had a negative feeling and $2\%$ had a positive feeling. Concerning Bernie Sanders, $83\%$ of the authors were neutral about him, $15\%$ had a negative feeling and $2\%$ had a positive feeling. As for Mike Pence, $85\%$ of the authors were neutral about him, $13\%$ had a negative feeling and $2\%$ had a positive feeling. Concerning Boris Johnson, $89\%$ of the authors were neutral about him, $9\%$ had a negative feeling and $2\%$ had a positive feeling. Moving to Donald Trump, $77\%$ of the authors were neutral about him, $18\%$ had a negative feeling and $5\%$ had a positive feeling. Overall, we conclude from this Figure that the most controversial (having higher positive and negative sentiments toward them) personages were Joe Biden and Donald Trump who were in a fierce competition for the $2020$ United States presidential election. This also hints that the COVID-$19$ pandemic had an effect on the people's general opinion regarding candidates in the $2020$ United States presidential election.

\section{Conclusion}
\label{sec:Conclusion}
We analyze in this work a dataset that contains news/message/boards/blogs in English about
COVID-$19$ for the period December $2019$ to March $2020$ from several social media platforms such as Facebook, LinkedIn, Pinterest, StumbleUpon and VK. Our results suggest that (1) the number of posts related to COVID-$19$ increased exponentially from December $2019$ to March $2020$; (2) interests of social media users changed from being health-oriented in December $2019$ to being economics-oriented in January $2020$, and news-oriented in February and March $2020$; (3) the amount of fake news increased six times from December $2019$ to March $2020$; (4) most of the news, including the illegitimate ones, originated from the United States; (5) people mostly had a neutral sentiment toward public figures with negative sentiments prevailing positive ones; (6) the most controversial public figures with more positive and negative sentiments in the studied period were Joe Biden and Donald Trump.

\section{Acknowledgment}
This work is partially funded by the Natural Sciences and Engineering Research Council of Canada (NSERC) under grant number RGPIN-$2020$-$04707$ and by the Universit\'{e} du Qu\'{e}bec en Outaouais (UQO).
\bibliographystyle{elsarticle-harv}
\bibliography{COVID_Detection_Archive}
\end{document}